\documentclass[conference]{IEEEtran}
\IEEEoverridecommandlockouts
\usepackage{cite}
\usepackage{amsmath,amssymb,amsfonts}
\usepackage{algorithmic}
\usepackage{graphicx}
\usepackage{textcomp}
\usepackage{xcolor}
\usepackage{verbatim}
\def\BibTeX{{\rm B\kern-.05em{\sc i\kern-.025em b}\kern-.08em
    T\kern-.1667em\lower.7ex\hbox{E}\kern-.125emX}}
\begin{document}

\title{The Petascale DTN Project: High Performance Data Transfer for HPC Facilities\\
\thanks{This work was supported by the Director, Office of Science, Office of Advanced Scientific Computing Research (ASCR), of the U.S. Department of Energy under Contract No. DE-AC02-05CH11231.

This manuscript has been authored by an author at Lawrence Berkeley National Laboratory under Contract No. DE-AC02-05CH11231 with the U.S. Department of Energy. The U.S. Government retains, and the publisher, by accepting the article for publication, acknowledges, that the U.S. Government retains a non-exclusive, paid-up, irrevocable, world-wide license to publish or reproduce the published form of this manuscript, or allow others to do so, for U.S. Government purposes.

Disclaimer: This document was prepared as an account of work sponsored by the United States Government. While this document is believed to contain correct information, neither the United States Government nor any agency thereof, nor the Regents of the University of California, nor any of their employees, makes any warranty, express or implied, or assumes any legal responsibility for the accuracy, completeness, or usefulness of any information, apparatus, product, or process disclosed, or represents that its use would not infringe privately owned rights. Reference herein to any specific commercial product, process, or service by its trade name, trademark, manufacturer, or otherwise, does not necessarily constitute or imply its endorsement, recommendation, or favoring by the United States Government or any agency thereof, or the Regents of the University of California. The views and opinions of authors expressed herein do not necessarily state or reflect those of the United States Government or any agency thereof or the Regents of the University of California.
}
}


\makeatletter
\newcommand{\linebreakand}{%
  \end{@IEEEauthorhalign}
  \hfill\mbox{}\par
  \mbox{}\hfill\begin{@IEEEauthorhalign}
}
\makeatother


\author{\IEEEauthorblockN{Eli Dart}
\IEEEauthorblockA{\textit{Lawrence Berkeley National Laboratory}\\
}
\and
\IEEEauthorblockN{William Allcock}
\IEEEauthorblockA{\textit{Argonne National Laboratory} \\
}
\and
\IEEEauthorblockN{Wahid Bhimji}
\IEEEauthorblockA{\textit{Lawrence Berkeley National Laboratory} \\
}
\and
\IEEEauthorblockN{Tim Boerner}
\IEEEauthorblockA{\textit{National Center for Supercomputing Applications} \\
}
\and
\IEEEauthorblockN{Ravinderjeet Cheema}
\IEEEauthorblockA{\textit{Lawrence Berkeley National Laboratory} \\
}
\and
\IEEEauthorblockN{Andrew Cherry}
\IEEEauthorblockA{\textit{Argonne National Laboratory} \\
}
\and
\IEEEauthorblockN{Brent Draney}
\IEEEauthorblockA{\textit{Lawrence Berkeley National Laboratory} \\
}
\and
\IEEEauthorblockN{Salman Habib}
\IEEEauthorblockA{\textit{Argonne National Laboratory} \\
}
\and
\IEEEauthorblockN{Damian Hazen}
\IEEEauthorblockA{\textit{Lawrence Berkeley National Laboratory} \\
}
\and
\IEEEauthorblockN{Jason Hill}
\IEEEauthorblockA{\textit{Oak Ridge National Laboratory} \\
}
\and
\IEEEauthorblockN{Matt Kollross}
\IEEEauthorblockA{\textit{National Center for Supercomputing Applications} \\
}
\and
\IEEEauthorblockN{Suzanne Parete-Koon}
\IEEEauthorblockA{\textit{Oak Ridge National Laboratory} \\
}
\and
\IEEEauthorblockN{Daniel Pelfrey}
\IEEEauthorblockA{\textit{Oak Ridge National Laboratory} \\
}
\and
\IEEEauthorblockN{Adrian Pope}
\IEEEauthorblockA{\textit{Argonne National Laboratory} \\
}
\and
\IEEEauthorblockN{Jeff Porter}
\IEEEauthorblockA{\textit{Lawrence Berkeley National Laboratory} \\
}
\linebreakand
\IEEEauthorblockN{David Wheeler}
\IEEEauthorblockA{\textit{National Center for Supercomputing Applications} \\
}
}
\maketitle
\begin{abstract}
The movement of large-scale (tens of Terabytes and larger) data sets between high performance computing (HPC) facilities is an important and increasingly critical capability. A growing number of scientific collaborations rely on HPC facilities for tasks which either require large-scale data sets as input or produce large-scale data sets as output. In order to enable the transfer of these data sets as needed by the scientific community, HPC facilities must design and deploy the appropriate data transfer capabilities to allow users to do data placement at scale.

This paper describes the Petascale DTN Project, an effort undertaken by four HPC facilities, which succeeded in achieving routine data transfer rates of over 1PB/week between the facilities. We describe the design and configuration of the Data Transfer Node (DTN) clusters used for large-scale data transfers at these facilities, the software tools used, and the performance tuning that enabled this capability.
\end{abstract}


\section{Introduction} \label{introduction}
While HPC facilities are well-known for capability-class machines that run large-scale simulations, data-intensive workloads are becoming increasingly common\cite{bib:Pivot2Data}. Though science collaborations make every effort to move computation to data whenever possible for efficiency reasons, sometimes data must be moved to computation. There are several reasons for this, including the assignment of computing allocations at multiple facilities to one project, the need to assemble a data set from multiple sources at one location for analysis, and the availability of unique resources at a location different from the current data storage location. In most cases, the movement of data to or between facilities happens at the direction of a human - this means that the attention of a person is being spent managing the transfer of the data rather than on the analysis of the data. While the transfer of the data is a critical component of the overall scientific workflow, the value and insight is realized in the analysis of the data. As data sets increase in scale over time, the human effort associated with transferring and assembling data sets also increases unless the tools used to move the data are scalable. In the current environment of exponential data growth, the availability of scalable data transfer methods often determines whether a given data analysis task is feasible at all.

The genesis of the Petascale DTN Project was a talk which described the use of multiple DOE HPC facilities (ALCF\cite{bib:ALCF}, NERSC\cite{bib:NERSC}, and OLCF\cite{bib:OLCF}) for running the HACC\cite{bib:HACC} code to do cosmological simulations and for analyzing the resultant data products. Moving multi-Terabyte data sets between facilities was challenging and consumed significant human effort, even though the facilities were all interconnected by the 100Gbps ESnet\cite{bib:ESnet} network. A project team was formed to generalize a solution to the data movement problem that would benefit not only the HACC team but users of the HPC facilities generally. Because DOE does not exist in isolation, the NCSA\cite{bib:NCSA} Blue Waters system was included in the project to benefit science teams that use both DOE- and NSF-funded facilities.

The goal of the project was to achieve routine data transfer performance at the level required to move 1 petabyte of data in a week, using a data transfer tool available in production at all the facilities, without the need for significant human effort on the part of facility users. In round numbers, moving 1PB per week requires about 13.25Gbps of throughput over the course of a week - the project chose a goal of 15Gbps to allow for headroom. A desired beneficial side effect was to create an existence proof that persistent, routine, high-speed data transfer was possible using production systems, and to create a body of best practice for configuring and deploying high-speed data transfer capabilities in support of data-intensive science and high performance computing.

The rest of the paper is organized as follows: Section \ref{dtn-interface} describes the location of data transfer nodes (DTNs) in the HPC facility network; Section \ref{dtnarch} covers design aspects of the DTN hosts; Section \ref{fs-interaction} describes the interaction between the DTNs and parallel file systems; Section \ref{dtn-tools} describes the data transfer software; Section \ref{perf-results} describes performance tuning considerations; Section \ref{related-work} describes other similar projects; Section \ref{future-work} touches on future work; and Section \ref{conclusions} concludes the paper.

\section{The DTN as HPC Facility External Data Interface} \label{dtn-interface}
Data transfers do not occur between file systems directly. Rather, they take place between servers which mount those file systems (and between data transfer applications running on those servers). In the Science DMZ\cite{bib:ScienceDMZ} model, these servers are called Data Transfer Nodes or DTNs. In the context of a HPC facility, it is important from a design perspective to think of the DTN as the external-facing interface to the parallel file system which is also mounted on the capability-class HPC resources at the facility. The DTN’s job in this context is to read or write data sets on the filesystem and move them to or from the network at high speed under the control of a properly-authenticated user or workflow, and the clear articulation of this role allows it to be generalized as part of HPC facility design. 

This configuration also avoids the disadvantages of a multi-copy workflow. There is no user data storage within the DTN servers in this configuration - all user data is stored on a separate filesystem mounted by the DTNs over a storage fabric. This keeps the large-scale data objects (the scientific user data) on the large parallel filesystem where it belongs, and where it is accessible to the other computing elements in the HPC facility. It also avoids the need to manage local storage on the DTN servers for user data transfers, and removes the limitations on storage system scalability imposed by fitting the storage into a single DTN server. 

The only systems that need to be configured for transferring files over long distance networks are the DTNs - the rest of the file system infrastructure can be focused on the internal aspects of the HPC facility. This is a key point of leverage in both computing facility and system design. The DTN runs the data transfer application, and together with the DTN servers this application moves the data between high-speed storage at the communicating HPC facilities via the network which interconnects them. Thus, the DTNs serve to integrate the HPC facility with the outside world from a high-speed data perspective. Figure \ref{fig:hpc-science-dmz}, adapted from the HPC facility diagram in \cite{bib:ScienceDMZ}, shows this integration - long distance data transfers are not made to the high performance parallel filesystem directly, but rather occur on the DTNs which mount the parallel file system. Also, the supercomputer nodes themselves do not need to be tuned or configured for long distance data transfer - the DTN cluster is explicitly configured for this purpose, relieving the supercomputer nodes of this added configuration complexity.

The concept of DTN specialization can be carried further as well, especially when different HPC facility functions require different configurations, different behaviors, or different security postures from the DTNs. Depending on facility needs and policies, it may be desirable to configure one cluster of DTNs with login access by users (e.g. to support legacy command line tools), and a different cluster for transfers controlled by means of a scalable API. Of the facilities in the project, the OLCF has the greatest degree of DTN specialization, with separate DTN clusters for interactive tools (e.g. SCP\cite{bib:ssh}, rsync\cite{bib:rsync}), Globus\cite{bib:Globus} use, and HPSS\cite{bib:HPSS} access, with an additional cluster of DTNs reserved for and controlled by the batch system scheduler. The batch scheduled DTNs are primarily useful within the facility and were critical to the effort of migrating data between the center-wide file systems for Titan \cite{bib:titan} and Summit  \cite{bib:summit}. This is a key point, in that the capabilities DTNs provide are useful and productive both for scientific users and for the operational teams which run the HPC facilities.

\begin{figure}[htbp]
\centerline{\includegraphics[width=0.5\textwidth]{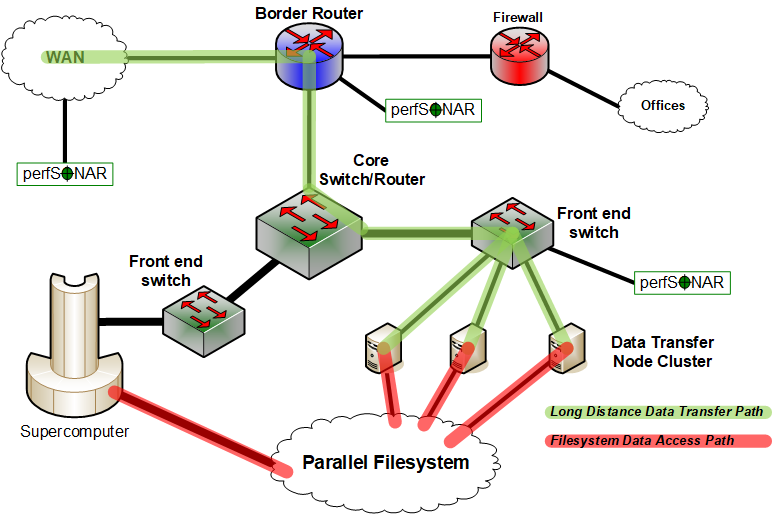}}
\caption{Simplified abstract network diagram of a HPC facility, showing the DTN cluster deployed in keeping with the Science DMZ model. The long distance data transfer and local file system access paths are highlighted, each using the appropriate DTN network interface.}
\label{fig:hpc-science-dmz}
\end{figure}

\section{DTN System Architecture} \label{dtnarch}
The DTN must do several things simultaneously. It must run the data transfer application, but it also must connect a fundamentally local technology (the internals of a parallel file system) to a fundamentally non-local technology (the wide area network), while maintaining consistently high performance levels. Several aspects of DTN server configuration became clear in the context of the project.

In order to connect to the wide area network, the DTN is essentially required to have an Ethernet\cite{bib:Ethernet} interface. Ethernet is the dominant network interface technology for Internet-connected servers, and high-speed Ethernet interfaces are readily available and well-understood. However, Ethernet is not necessarily used for the internals of a parallel file system or supercomputer interconnect. In many cases, InfiniBand\cite{bib:InfiniBand} is used for these internal networks. Therefore, the DTN must connect to both, reading data from one network technology and writing it to the other. This gives HPC facility DTNs multiple network interfaces in the general case, and can contribute to system complexity. In addition, operational teams sometimes desire interface redundancy, adding connections for failover and reliability. This can further increase system complexity.

While there are some theoretical advantages to powerful and complex DTNs (e.g. extra Ethernet interfaces for redundancy/failover or added performance), during the project simplicity emerged as a key contributor to consistently high performance. At the beginning of the project, some of the facilities had larger sets of simple DTNs used for data transfer, and some had smaller sets of powerful and complex DTNs with significant internal redundancy. Achieving consistent performance was straightforward with a simple DTN that had a single Ethernet interface and a single InfiniBand interface, while a complex DTN with multiple Ethernet interfaces and multiple InfiniBand interfaces had significantly varied performance. Eventually, all the participating facilities adopted the simple DTN model, with the data transfer tool spreading the load across multiple DTNs. The DTN-level parallelism available in advanced data transfer tools is a critical point of scalability, because HPC facilities are able to add nodes as needed to increase performance. This makes the data transfer tools that can simultaneously use multiple DTNs highly scalable - a key capability for managing data growth without substantially increasing human effort.

Another key contributor to a high performance DTN design is CPU clock rate. Packet processing is an inherently serial task\cite{bib:Partridge}, and while modern data transfer tools use multiple data connections in parallel a DTN must manage high packet and data rates on its network interfaces. In general, a smaller number of cores with a higher clock rate is preferable, but within reason. This is an element of system design that requires careful consideration by the operational team which designs and deploys the DTN cluster - different workloads will require different configurations. However, it is generally the case that sacrificing too much clock speed for additional cores is not a good DTN design trade-off. If more parallel threads are needed it is better to deploy additional DTNs with higher-speed CPUs. Additional DTNs have the added advantage of increasing the amount of RAM available for buffering and increasing the number of file system mount points - these are important scalability considerations as well. Finally, if integrity checksums are used (as they should be for large-scale data transfers\cite{bib:checksum-disagree}) then each data transfer job is also a compute job from the perspective of the DTN\cite{bib:integrity} - this also argues in favor of higher CPU clock rate so that checksum verification can keep up with I/O.

In considering how the DTNs connect to the facility network, it is important to consider both the speed of the DTN Ethernet interface and the buffering in the switches or routers that connect directly to the DTN, as well as the speed of other DTNs that exist at remote sites. If there is a mismatch between the Ethernet interface speed on the local DTN and the Ethernet interface speed on the remote DTN, the side with the slower Ethernet interface will need to have more buffering in the router or switch at the bottleneck link in order to avoid packet loss from the speed mismatch. Dropped packets cause performance problems for TCP as described in \cite{bib:ScienceDMZ} and preventing packet loss in production operation is a key aspect of network and system design for DTNs.

\section{Interaction With The File System} \label{fs-interaction}
 Each facility has the potential for deploying different filesystem technologies, and some facilities have multiple large-scale parallel filesystems, which may be based on different technologies for a variety of reasons. There are two primary parallel filesystem technologies deployed in the HPC facilities in the project - they are the IBM Spectrum Scale \cite{bib:gpfs} and Lustre \cite{bib:lustre} filesystems. Each have client side tuning (done on the DTN) that is important. As noted in Section \ref{dtnarch} the DTN requires multiple interfaces - Ethernet to the Internet and primarily Infiniband to the parallel file system. Tuning of each network interface is handled separately, though TCP/IP stack parameters are typically global. Additional tuning is required for the parallel file system client and potentially system kernel modules. It should be noted that there is no single "right" way to do this, and the HPC facility storage team has to make engineering tradeoffs given the filesystem configuration and the resources of the DTNs.

\subsection{Lustre Tuning}
The Lustre parallel file system has different levels of tunables. The HPC facilities have expertise in this area, and the tunables are specific to the configuration and architecture of the facility's deployment. A high level overview would be to perform any LNET tuning and then Lustre tuneables for the client. In general the tuning should allow the greatest amount of traffic across the interconnect, provide sufficient caching, and not hinder metadata operations while keeping within a reasonable memory footprint for the available memory in the DTN. In particular, performance benefits from increasing the \verb peer_credits  parameter to the maximum available for the underlying network interface to the Lustre file system as well as \verb max_rpcs_in_flight  and \verb max_mod_rpcs_in_flight  to the greatest extent possible given the system memory on the DTN. The HPC facilities experts have each explored this parameter space for their site and have developed configurations that allow for the best possible performance for the DTNs. Examples of tunables can be found in the Lustre Operations Manual \cite{bib:lustreops}.

The Lustre configuration appropriate for one facility may not be the same as the appropriate Lustre configuration in a different environment. Unfortunately, this is a case where the team running the production storage system and DTN infrastructure must make a set of engineering decisions, often based on experimentation. Two examples follow, illustrating the commands (as they might be deployed in an rc.local script) to set three key tuning parameters used by two different HPC facilities.
\begin{verbatim}
lctl set_param osc.*.max_rpcs_in_flight=32
lctl set_param mdc.*.max_rpcs_in_flight=64
lctl set_param llite.*.statahead_max=8192
      
      
lctl set_param osc.*.max_rpcs_in_flight=16
lctl set_param mdc.*.max_rpcs_in_flight=8
lctl set_param llite.*.statahead_max=32

\end{verbatim}
The point of this comparison is not to provide specific examples for use by others, but rather to illustrate the need for the storage team to understand the local environment, and make decisions based on testing real-world scenarios relevant to the workload of that HPC facility.

\subsection{Spectrum Scale/GPFS Tuning}
With IBM's Spectrum Scale file system, the parameters for client performance are less visible than Lustre due to the licensing and support models provided by IBM. A cursory look through the Parameters for performance tuning and optimization guide on IBM's Knowledge Center \cite{bib:ibmkc} provides a small list of what is available. Performance tuning should be performed based on the available resources of the file system, interconnect, and DTN to allow the greatest level of caching and throughput performance. If the DTN is using Infiniband to connect to the Spectrum Scale file system setting \verb RUN_MLNX_TUNE=yes  will setup irq affinity for the NICs and Infiniband interfaces. Setting \verb maxMBpS  to a high enough value supported by system memory increases the client's performance. It is important to ensure that \verb maxblocksize  is matched with the file system cluster. An additional tuning parameter is \verb maxFilesToCache  which allows the client to cache files and does not impact the pagepool but does consume system memory. There is a tuning parameter for increasing the number of threads that the Spectrum Scale client uses, and setting \verb ignorePrefetchLUNCount  \verb yes  ensures that the threads initiated isn't dependent on the number of LUNs in the Spectrum Scale file system.

\section{Data Transfer Software} \label{dtn-tools}
For many users and collaborations, rsync has been the data transfer tool of choice for many years. While rsync typically transfers data correctly, it is typically slow -- this is both because rsync is single-threaded and therefore unable to scale beyond a single DTN, and because it runs over SSH\cite{bib:ssh} in most cases. While high performance patches\cite{bib:hpn-ssh} exist which fix some of the limitations of SSH, rsync over SSH has inherent scalability limitations, especially if human effort is factored in. Because of these dual limitations, rsync was not considered an adequate tool. Also, SCP (which also runs over SSH), is subject to the same performance limitations as rsync over SSH. These are considered legacy tools, and are generally not suitable for modern high-speed data transfer workflows.

There was one data transfer tool that was already installed at all four facilities, had high performance, and had the ease of use to scale up data transfer performance without similarly scaling human effort -- the project chose Globus. Globus is able to manage multiple DTNs and distribute one transfer across multiple DTNs (as long as the data set contains a sufficiently large number of files), has an easy-to-use user interface, and also ensures the correctness of data transfer through the use of integrity checksums. The existence of Globus deployment at the HPC facilities also allows for the easy assembly of data sets published on data portals using the Modern Research Data Portal (MRDP)\cite{bib:MRDP} design pattern, such as the NCAR/UCAR Research Data Archive\cite{bib:RDA}, which also use Globus. A variety of other data transfer tools and platforms also exist in the broader scientific community - an in-depth evaluation of several such tools is provided by \cite{bib:tool-survey}.

\section{Performance Tuning and Test Results} \label{perf-results}
The project team used a broad range of performance tuning and troubleshooting techniques over the course of the project. Regular testing, both using the DTNs themselves and testing the networks with perfSONAR\cite{bib:perfsonar}, proved invaluable. Also key was the ability to test between multiple sites and therefore "triangulate" to determine the likely location of issues impeding performance. The ability to work collaboratively to identify the organization that was likely to own the misbehaving component (HPC facility, local network, wide area network), and then understand what technology was causing the problem (network, storage system, DTN, etc.) is crucial for effective troubleshooting, and regular test and measurement is critical for spotting problems early. Network testing from site to site tells whether the network path is capable of supporting high speed transfers, DTN filesystem mount testing indicates whether the DTN itself can read or write data at the required speeds, and so forth.

It is important to note that testing individual components of a data transfer workflow (filesystem performance on the DTN, wide area network speeds using perfSONAR, wide area network speeds using perfSONAR tools on the DTNs, etc) is necessary but not sufficient. One of the indications that system complexity was problematic was a case where individual subsystem tests performed at or above the target rate, but the end to end data transfer rate was sub-par. Only after adopting a simpler DTN design at that facility did end to end performance improve.

The facilities made several enhancements to their networks and DTN clusters over the course of the project. By the end of the project, all the sites had 100Gbps network connections between the DTN clusters and the wide area network. In addition, several sites increased the number of DTNs available, and the simplified DTN design of one or two Ethernet interfaces and minimal InfiniBand interfaces was adopted by all sites for performance consistency.

The DTN hosts themselves need to be configured to support high performance long-distance data transfers. The information to do this is readily available from knowledge bases such as the ESnet Fasterdata Knowledgebase\cite{bib:Fasterdata}, but this must be done or the DTN’s network configuration will limit its capability. For example, the DTNs are configured by default to use kernel buffer autotuning for TCP connections, but the default parameters are insufficient for high-speed long distance data transfers and so must be increased for DTNs. High speed processors, fast network interfaces, and sufficient memory are important, but if the DTN is not configured for wide area data transfers it will still perform poorly in the general case. The host configuration details are published as a separate artifact\cite{bib:test-dataset}.

In addition to the hosts, networks, and filesystems, Globus has performance tuning parameters as well. These are covered under the "network use" configuration\cite{bib:globus-tuning}. For large, high-speed DTN cluster installations, the "Aggressive" setting achieves the highest performance of the standard option set (which are "Minimal," Default," and "Aggressive"). Some of the sites in the project use custom Globus tuning settings, but they are similar to the "Aggressive" settings.

One unforeseen change provided an unexpected increase in performance: the Globus team changed the job scheduling algorithm for transfer jobs containing large numbers of files. The batch size for transfers containing large numbers of files was increased from 1,000 to 10,000 if if the first batch transferred successfully. This resulted in significantly increased performance, presumably because the DTNs spent less time waiting for batches of files to finish transferring due to the smaller number of batches for a given transfer.

In the end, multiple aspects of the full data transfer workflows had to be addressed. DTN configuration, Globus configuration, and network configuration all played important roles. No single parameter change was solely responsible for high performance - all components have to be in good working order and working well together in order to perform consistently well.

The data transfer tests were performed by transferring a single directory containing a subset of the data output from a HACC simulation run\cite{bib:test-dataset}. The data transfers were performed using the Globus GUI, which works like a normal graphical file transfer tool, just as any normal scientific user would use the Globus GUI to transfer a data set. This was a deliberate decision - the intent was to measure the real-world behavior of production systems during normal use. The data set directory contained many subdirectories, and a range of file sizes. In all, the data set consisted of 211 directories and 19,260 files, ranging in size from zero bytes to 11.3GB. The total data set size was 4442781786482 bytes (about 4.4TB).

\begin{table}[h]
\caption{File size composition of test data set}
\centering
\begin{tabular}{|c|c|}
\hline
\multicolumn{2}{|l|}{\textbf{File count per order of magnitude}}  \\ \hline
\textbf{\textit{Size (bytes)}}  & \textbf{\textit{Number of Files}}    \\ \hline 
0 - 10                        & 7                               \\ \hline
10 - 100                      & 1                               \\ \hline
100 - 1K                      & 59                              \\ \hline
1K - 10K                      & 3170                            \\ \hline
10K - 100K                    & 1560                            \\ \hline
100K -1M                      & 2817                            \\ \hline
1M - 10M                      & 3901                            \\ \hline
10M - 100M                    & 3800                            \\ \hline
100M -1G                      & 2295                            \\ \hline
1G - 10G                      & 1647                            \\ \hline
10G - 100G                    & 3                               \\ \hline
\end{tabular}

\end{table}

Test transfers were performed on production systems during normal operation - this was a deliberate decision made with the intent of measuring the real-world circumstances that a normal DTN user would encounter during routine data transfers. Table \ref{table:starting-rates} shows the data transfer rate between DTN clusters at the beginning of the project.

\begin{table}[htbp]
\caption{Pairwise Data Transfer Rate at Project Start, Gigabits Per Second}
\begin{center}
\begin{tabular}{|c|c|c|c|c|}
\hline
\textbf{ }&\multicolumn{4}{|c|}{\textbf{Destination Site}} \\
\cline{1-5} 
\textbf{Source Site} & \textbf{\textit{ALCF}} & \textbf{\textit{NCSA}} & \textbf{\textit{NERSC}} & \textbf{\textit{OLCF}} \\
\hline
\textbf{\textit{ALCF}}  &       & 13.4  & 10.0  & 10.5  \\  \hline
\textbf{\textit{NCSA}}  & 8.2   &       & 6.8   & 6.9   \\  \hline
\textbf{\textit{NERSC}} & 7.3   & 7.6   &       & 6.0   \\  \hline
\textbf{\textit{OLCF}}  & 11.1  & 13.3  & 6.7   &       \\  \hline
\end{tabular}
\label{table:starting-rates}
\end{center}
\end{table}

In the end, data transfers between all combinations of endpoints significantly exceeded the project goal of 15Gbps. Table \ref{table:end-rates} compares the data transfer rates at the beginning and at the end of the project.

\begin{table}[htbp]
\caption{Pairwise Average Data Transfer Rate at Project End, Gigabits Per Second}
\begin{center}
\begin{tabular}{|c|c|c|c|c|}
\hline
\textbf{ }&\multicolumn{4}{|c|}{\textbf{Destination Site}} \\
\cline{1-5} 
\textbf{Source Site} & \textbf{\textit{ALCF}} & \textbf{\textit{NCSA}} & \textbf{\textit{NERSC}} & \textbf{\textit{OLCF}} \\
\hline
\textbf{\textit{ALCF}}  &       & 50.0  & 35.0  & 46.8  \\  \hline
\textbf{\textit{NCSA}}  & 56.7  &       & 22.6  & 34.7  \\  \hline
\textbf{\textit{NERSC}} & 42.2  & 33.7  &       & 39.0  \\  \hline
\textbf{\textit{OLCF}}  & 47.5  & 43.4  & 33.1  &       \\  \hline

\end{tabular}
\label{table:end-rates}
\end{center}
\end{table}

Fig. \ref{fig:dtn-performance} summarizes the performance levels over multiple test runs at the end of the project.

\begin{figure}[htbp]
\centerline{\includegraphics[width=0.5\textwidth]{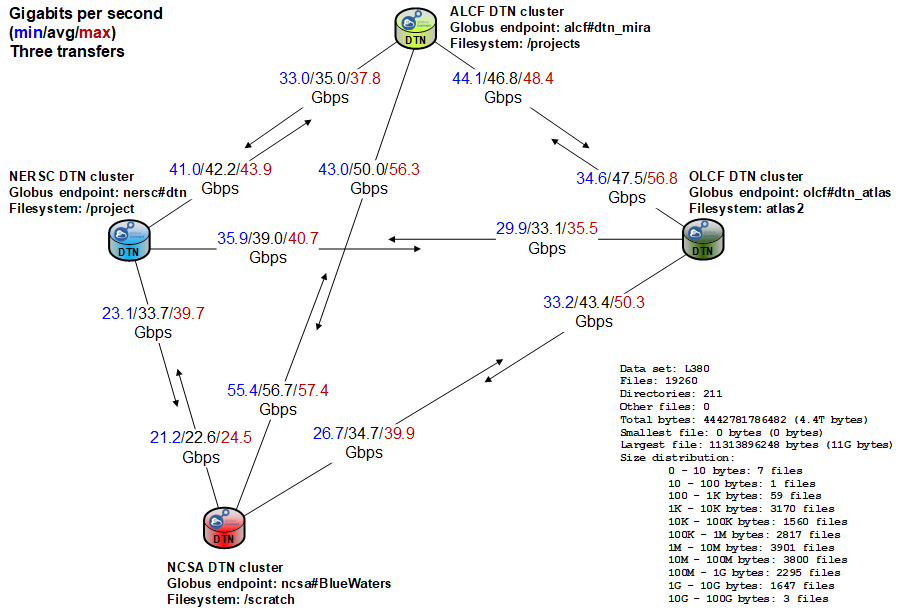}}
\caption{Performance between DTN clusters at the conclusion of the project. The composition of the data set indicates the large file count and variation in file sizes of the reference data set.}
\label{fig:dtn-performance}
\end{figure}

\section{Related Work} \label{related-work}
The interconnection of high-speed computing facilities has been done in the past - for example, the TeraGrid project\cite{bib:teragrid} and its successor XSEDE\cite{bib:xsede} have provided dedicated high-speed networks to interconnect HPC facilities, though XSEDE has since transitioned to a dedicated virtual network on shared physical infrastructure. Also, the ASC DisCom network interconnected computing resources for the DOE's stockpile stewardship program\cite{bib:discom}. These efforts have several aspects in common with the Petascale DTN Project (e.g. connecting HPC sites or facilities together for high speed data transfer), but the Petascale DTN Project is different in that we explicitly use a shared, multi-user (i.e. non-dedicated) science network for the data transfers.

\section{Future Work} \label{future-work}
The Petascale DTN Project has been successful, both in demonstrating that high-speed data transfer is a capability that is deployable in production on multi-user systems and in actually deploying that capability for routine use by many thousands of scientific users across four computing facilities interconnected by production multi-user networks. However, this project has also provided a foundation for future work in the scientific community to make data sets available for analysis using national HPC resources.

Now that it is possible to move data to and between the four facilities at high speed, the next step is to expand the capability further within the scientific community. There are three primary classes of data resources that would benefit from enhancements using the best practices described here: site and campus clusters, data portals, and experimental or observational data sources.

Computing clusters at labs, universities, departments, and other similar entities are typically similar in architecture to a national HPC facility, but at smaller scale. The same design pattern, consisting of a well-configured DTN cluster with good design and good tools as the external interface to a large filesystem, works well in a smaller environment just as it does in a national HPC facility. The advantage of enhancing the computing clusters in the broader community with DTN clusters is clear - not only would they be able to transfer data well between them, but they would integrate well with national HPC facilities.

Data portals are large repositories of scientific data, much of which could be beneficially analyzed at HPC facilities using modern data analysis techniques. The Modern Research Data Portal design pattern (MRDP) \cite{bib:MRDP} describes a design similar to the one presented here in that a DTN cluster is used as the external-facing interface for the data store. When a MRDP-enhanced data portal provides a user with a link to a data set, that link consists of a Globus job instead of a HTTP link to a file. This provides the data portal with the same scalability that HPC facilities get from deploying DTN clusters with a high-performance data transfer tool such as Globus. Data portals which have been enhanced by adopting the MRDP design pattern integrate well with Petascale DTN Project facilities for data transfer, and expanding the deployment of the MRDP design pattern using DTN clusters would increase the ability of the scientific community to analyze the data in those portals at scale.

Finally, there is an increasing need to integrate HPC facilities with experimental and observational facilities \cite{bib:eod-report}. The deployment of DTN clusters at experimental and observational facilities would provide a scalable platform for transferring data to HPC facilities for analysis.

\section{Conclusions} \label{conclusions}
Several key elements of best practice can be derived from this work, as described below.

For large HPC facilities it is incredibly valuable to deploy a cluster of DTNs in a way that allows a single scalable data transfer tool to manage transfers across multiple DTNs in the cluster. By explicitly incorporating parallel capabilities (e.g. the simultaneous use of multiple DTNs for a single job)  into the design, the external interface to the HPC facility storage system can be scaled up as needed by adding additional DTNs, without changing the underlying tool. This combination of a scalable tool and a scalable architecture is critical as the HPC community moves towards the Exascale era while at the same time increasing the number of research projects that have large-scale data analysis needs. However, this design pattern applies equally well to institutional or departmental DTN clusters (e.g. at universities) - it allows for easy scalability as needs grow, and enables high-speed integration with national facilities if both sides deploy interoperable tools.

In building a cluster of DTNs, it is important to keep the design of the individual DTNs simple so that their behavior is consistent. All the facilities converged on a design consisting of a cluster of many simple DTNs with very similar external network interface specifications rather than a small number of highly-redundant and complex DTNs.

Deploying a suite of performance monitoring tools that record the performance at regular intervals is key to determining both a performance baseline of a particular set of DTNs and the impact of configuration changes on the overall deployment. Utilizing a data visualization and aggregation tool can allow teams to create views of performance based on the gathered monitoring data.

Tuning individual components (e.g. filesystem mount performance, long-distance network performance) is necessary but not sufficient. Ultimately everything must work together -- filesystem, DTNs, network, and data transfer tool -- in a way that consistently achieves high performance for the user community, in production operation, without constant troubleshooting. 

Finally, the data transfer tool itself is of paramount importance. While it is critical to have a system design that can scale to the levels required to meet the science mission, human time and attention are precious resources -- the tools presented to users must make them more productive, not less. The ability to easily initiate a transfer and then allow the tools to manage the transfer without direct human intervention is incredibly valuable, and is a key contributor to the scale at which scientists will be able to do data analysis as data sets grow ever larger.

\section*{Acknowledgments}
The project team would like to acknowledge the contributions of Jason Anderson of the OLCF, who lost his battle with cancer in August 2019. It is our hope that the HPC facility enhancements he helped realize will one day help to cure cancer for everyone.

This research used resources of the Argonne Leadership Computing Facility, which is a U.S. Department of Energy Office of Science User Facility operated under contract DE-AC02-06CH11357. 

Argonne National Laboratory's work was supported by the U.S. Department of Energy, Office of Science, under contract DE-AC02-06CH11357.

This research used resources of the National Energy Research Scientific Computing Center (NERSC), a U.S. Department of Energy Office of Science User Facility operated under Contract No. DE-AC02-05CH11231.

This research used resources of the Oak Ridge Leadership Computing Facility at the Oak Ridge National Laboratory, which is supported by the Office of Science of the U.S. Department of Energy under Contract No. DE-AC05-00OR22725.

This research is part of the Blue Waters sustained-petascale computing project, which is supported by the National Science Foundation (awards OCI-0725070 and ACI-1238993) and the State of Illinois.




\vspace{12pt}

\end{document}